\documentclass[aps,prl,twocolumn,showpacs,superscriptaddress]{revtex4}
\usepackage{amsfonts}
\usepackage{amsmath}
\usepackage{amssymb}
\usepackage{graphicx}
\usepackage{epsfig}
\usepackage{bm}
\usepackage{mathrsfs}
\usepackage{color}
\usepackage{hyperref}

\def\bfr{{\bf r}}

\begin{document}

%\title{Newton-Schr\"odinger Equations are not derivable from General Relativity + Quantum Field Theory}
%\author{C. Anastopoulos}
%\affiliation{Department of Physics, University of Patras, 26500 Patras, Greece}
%\author{B. L. Hu}
%\affiliation{Maryland Center for Fundamental Physics and Joint Quantum Institute,\\
%University of Maryland, College Park, Maryland 20742, USA}

%\ead{anastop@physics.upatras.gr,blhu@umd.edu}
%Abstract: In this note we show that Newton-Schr\"odinger Equations (NSEs) do not follow from general relativity (GR) and quantum field theory (QFT) by way of two considerations: 1) Taking the nonrelativistic limit of the semiclassical Einstein equation, the central equation of relativistic semiclassical gravity, a fully covariant theory based on GR+QFT with self-consistent backreaction of quantum matter on the spacetime dynamics; 2) Working out a model [see C. Anastopoulos and B. L. Hu, Class. Quant. Grav. 30, 165007 (2013)] with  a  matter scalar field interacting with weak gravity, in procedures analogous to the derivation of the nonrelativistic limit of quantum electrodynamics. We conclude that the coupling of classical gravity with quantum matter can only be via mean fields,  there are no $N$-particle NSEs and theories based on Newton-Schr\"odinger equations assume unknown physics.

%\date{Feb 2, 2014}
% \maketitle

{\bf Comments on Newton-Schr\"odinger Equations} pertaining to \cite{ChenNS} and relevant references therein.
The Newton-Schr\"odinger  equations (NSE) play a prominent role in alternative quantum theories (AQT)\cite{BassiRMP} and have served  as a theoretical platform for experimentalists to investigate the interaction of quantum matter with classical gravity. The latest development is the application of a many- body NSE to macroscopic mechanical objects \cite{ChenNS}. In this note we show that NSEs do not follow from general relativity (GR) and quantum field theory (QFT). We come to this conclusion from two considerations: 1) Working out a model (see \cite{AHmasteq}) with matter described by a  scalar field interacting with weak gravity, solve the constraint, canonically quantize the system then take the nonrelativistic (NR) limit. This procedure is the exact analogue of deriving the NR limit of quantum electrodynamics (QED). %Let's call this the QED route.
2) Taking the NR limit of the semiclassical Einstein equation (SCE), %where matter is described by a quantum field and its expectation value with respect with some quantum state acts as a source in driving the Einstein equation. Let's call this the SCE route. SCE Eq is
the central equation of relativistic semiclassical gravity (RSCG)\footnote{There are four levels of semiclassical gravity theories \cite{HuEmQM} and one needs be careful which level one refers to when debating issues, better use the  most developed level \cite{HuVerLivRev}.}, a fully covariant theory based on GR+QFT with self-consistent backreaction of quantum matter on the spacetime dynamics \cite{HuVerLivRev}.
We first lay out two key physical differences: %between theories based on NSE and those obtained from GR+QFT, such as via the two routes stated above.

A. In NSE the {\it gravitational self-energy}  defines non-linear terms in Schr\"odinger's equation.  In Diosi's theory \cite{Diosi}, the gravitational self-energy defines a stochastic term in the master equation.
With GR+QFT gravitational self-energy only contributes to mass renormalization, at least in the weak field (WF) limit.The Newtonian interaction term at the field level induces a divergent self-energy contribution to the single-particle Hamiltonian. It does not induce nonlinear terms to the Schr\"odinger equation for any number of particles.
%From the perspective of GR+QFT the NS equation {\em is not} the evolution equation for the single-particle wave function.\\

%B. A severe problem of the NSE when applied to a single-particle wave function is its {\it probabilistic interpretation}. Consider two statistical ensembles of particles one of which is described by the wave-function $\psi_1(\bfr)$ and the other by the wave function $\psi_2(\bfr)$. The ensemble obtained from mixing these ensembles with equal weight is described in standard quantum theory by the density matrix $\rho(\bfr, \bfr') = \frac{1}{2} [\psi_1(\bfr) \psi^*_1(\bfr') +   \psi_2(\bfr) \psi^*_2(\bfr')]$. The usual Schr\"odinger evolution guarantees that the probabilistic interpretation of the density matrix remains consistent under time evolution $\rho_t(\bfr,\bfr') = \frac{1}{2} [\psi_1(\bfr,t) \psi^*_1(\bfr',t) +  \frac{1}{2} [\psi_2(\bfr,t) \psi^*_2(\bfr',t)]$. This property does not apply for non-linear evolutions of the wave-functions.

B. The {\it  one-particle NS equation}
%(Note it is long known \cite{HarHor} that RSCG is the theory obtained as the large N limit of N component quantum fields living in a curved spacetime. )  There are important differences:
appears as the Hartree approximation for $N$ particle states as $N \rightarrow \infty$. Consider the ansatz $|\Psi\rangle = |\chi\rangle \otimes |\chi \rangle \ldots \otimes|\chi \rangle$ for a $N$-particle system. At the limit $N \rightarrow \infty$ the generation of particle correlations in time is suppressed and one gets an equation which reduces to the NS equation for $\chi$ \cite{Alicki,AdlerTDG}. However, in the Hartree approximation, $\chi({\bf r})$ is \textit{not} the wave-function $\psi(\bfr)$ of a single particle, but a \textit{collective variable} that describes a system of $N$ particles under a mean field approximation. %The NS equation is like the quantum analogue of the classical Vlasov equation which is obtained .

%In what follows we will show that the only meaningful description of quantum matter interacting with classical gravity is if the matter degrees of freedom are described in terms of quantum fields, \textit{not} in terms of single-particle wave functions in quantum mechanics.

%\subsection{NS Equation not from GR + QFT}

\paragraph{The NSE} governing the wave function of a single particle $\psi(\bfr,t)$ is of the form $i {\partial}\psi /{\partial t}  = - \frac{\hbar^2}{2m} \nabla^2 \psi + m^2 V_N[\psi]  \psi $ Eq.(1), %\label{NS} \end{eqnarray}
where %the potential energy $U(\bfr)$ from the gravitational interaction is of the form $U(\bfr)= m V_N(\bfr)$, where
$V_N(\bfr)$ is the (normalized) gravitational (Newtonian) potential given by $V_N({\bf r},t) = - \int d{\bf r'} |\psi({\bf r'},t)|^2 /|{\bf r} - {\bf r'}|.$ It satisfies the Poisson equation $\nabla^2 V_N = 4 \pi G \mu,$
with the mass density $\mu = m|\psi(\bfr,t)|^2 $ being the nonrelativistic  limit of energy density $\varepsilon = T_{00}$.% defined below. %To show where the problems are,  we first describe the SCE route because it appears to be closest to NSE but two fundamental differences led to the two points made above.

\paragraph{The semiclassical Einstein equation} is of the form
$G_{\mu \nu} = 8 \pi G \langle \Psi|\hat{T}_{\mu \nu}|\Psi \rangle$ Eq.(2), where $\langle \hat{T}_{\mu \nu} \rangle$ is the expectation value of the stress energy density operator $\hat{T}_{\mu \nu}$ with respect to some  quantum state $| \Psi \rangle$ of the field.  In the weak field limit the spacetime metric has the form $ds^2 = (1 - 2 V)dt^2 - d{\bf r}^2$, and the non-relativistic limit of the SCE Eq becomes $\nabla^2 V = 4 \pi G \langle \hat{\varepsilon}\rangle, $
where $\hat{\varepsilon} = \hat{T}_{00}$ is the energy density operator. This can be solved to yield
$ V(\bfr) = - G \int d{\bf r'} {\langle \hat \varepsilon({\bf r'})\rangle}/{|{\bf r} - {\bf r'}|}$ Eq.(3).  Note $\langle \hat \varepsilon(\bfr')\rangle$ in general contains ultraviolet divergences and need be regularized via known procedures.% (see, e.g., \cite{BirDav}).

%In the nonrelativistic limit we then get $V({\bfr}) = - G \int d \bfr' {\langle \Psi|\hat{\mu}_{reg}({\bfr'})|\Psi\rangle}/{|{\bfr} - {\bfr'}|},$ where $\hat{\mu}_{reg}$ is the regularized mass density operator. The evolution of the quantum field is described by the Hamiltonian %\begin{eqnarray}
%$\hat{H} = - \frac{1}{2m} \int d \bfr \hat{\psi}^{\dagger}({\bfr}) \nabla^2 \hat{\psi}({\bfr}) + G \int d \bfr d \bfr' \hat{\mu}_{reg}({\bfr}){\langle \Psi|\hat{\mu}_{reg}({\bfr'})|\Psi\rangle}/{|{\bfr} - {\bfr'}|}, $%\label{nrSCE}
%where $\hat\psi$ is the annihilation field operator. Projecting $|\Psi \rangle = | \chi \rangle$ into single-particle states, $| \chi \rangle = \int d{\bf r} \hat{\psi}^{\dagger}({\bf r}) \chi({\bf r}) | 0 \rangle,$ % \label{1par}
%one gets $\langle \Psi|\hat{\mu}_{reg}({\bfr})|\Psi \rangle = |\chi(\bfr)|^2$ %at the limit $\epsilon \rightarrow 0$.

Two key differences between the NR limit of SCE and NSE are: i) the energy density $\hat \varepsilon(\bfr)$ is an operator, not a c-number. The Newtonian potential is not a dynamical object in GR, %just like the electric potential is not dynamical in QED,
but subject to constraint conditions.  ii) the state $|\Psi\rangle$ of a field is a  $N$-particle wave function. Quantum matter is coupled to classical gravity as a mean-field theory, well defined only when $N$ is sufficiently large.

The (misplaced) procedure  leading one from SCE to a NS equation  is the treatment of $ m |\psi(\bfr,t)|^2 $ as a mass density for a single particle, while in fact the mass density is a quantum   operator $\hat{\epsilon}({\bf r}) = \hat{\psi}^{\dagger}({\bf r}) \hat{\psi}({\bf r})$ in the QFT Hilbert space. Not treating these quantities as operators bears the consequences A and B.

To cross check these observations we have carried out an independent calculation following the same procedures as in obtaining  the non-relativistic limit of QED. Treating the matter degrees of freedom  in terms of quantum fields $\hat{\psi}({\bf r})$ and $\hat{\psi}^{\dagger}({\bf r})$, we obtain the Schr\"odinger equation for the fields
$ i \partial |\Psi \rangle / \partial t = \hat{H} |\Psi \rangle,$
with %\begin{eqnarray}
$\hat{H} = - \frac{\hbar^2}{2m} \int d{\bf r} \hat{\psi}^{\dagger}({\bf r}) \nabla^2 \hat{\psi}(\bfr)
- G \int  \int d{\bf r} d{\bf r'} {(\hat{\psi}^{\dagger}\hat{\psi})({\bf r}) (\hat{\psi}^{\dagger}\hat{\psi})({\bf r'})}/{|{\bf r} - {\bf r'}|} $ Eq.(4). %\label{ham} \end{eqnarray}
%This equation can be obtained  \cite{AHmasteq}  from the theory of a scalar field interacting with gravity classically, then taking the non-relativistic limit and then quantizing the field. It is also widely employed in condensed matter systems, with a Coulomb potential for electrostatic interaction replacing the Newtonian potential for gravitational interaction (see \cite{AH-NS} for details).
This equation is very different from the NSE when considering a single particle state. For single-particle states the gravitational interaction leads only to a mass-renormalization term (similar to mass renormalization in QED). This is point A we made above. %The same mistake of treating a field as a product state of single particle wave function leads to the difference described in point B above.
Using the Hartree approximation to Eq. (4) leads to the same result as the NR WF limit of SCE, not NSE. (Point B).
Details are in \cite{AH-NS}.

Our analysis via two routes based on GR+QFT shows that NSEs are not derivable from them. Coupling of classical gravity with quantum matter can only be via mean fields. There are no $N$-particle NSEs. Theories based on Newton-Schr\"odinger equations assume unknown physics.
%and they do not describe the wave function of single particles, one or many.

C. Anastopoulos, {\it University of Patras, Greece} and B. L. Hu, {\it University of Maryland, USA}.  February 12, 2014.

\end{document}